\documentclass{acm_proc_article-sp}
\pdfoutput=1
\usepackage{graphicx} 
\usepackage{listings}

\newtheorem{definition}{Definition}

\def\spm{\mbox{SPM}}
 
\def\expwindow{\mbox{ExpW}}
\def\monwindow{\mbox{MonW}}

\def\malstone{MalStone } \def\malgen{MalGen }
\def\malstonens{MalStone} \def\malgenns{MalGen}

\begin{document}

\title{\malstonens: Towards A Benchmark for \protect\\ 
Analytics on Large Data Clouds}

\numberofauthors{5} \author{
  \alignauthor Collin Bennett \\
  \affaddr{Open Data Group}\\
  \affaddr{400 Lathrop Ave Suite 90}\\
  \affaddr{River Forest IL 60305}\\
 \alignauthor Robert L. Grossman\titlenote{This author is the
corresponding author.  He is
    also a faculty member at the University of Illinois at Chicago.} \\
  \affaddr{Open Data Group}\\
  \affaddr{400 Lathrop Ave Suite 90}\\
  \affaddr{River Forest IL 60305}\\
 \and 
  \alignauthor David Locke\\
  \affaddr{Open Data Group}\\
  \affaddr{400 Lathrop Ave Suite 90}\\
  \affaddr{River Forest IL 60305}\\
 \alignauthor Jonathan Seidman \\
  \affaddr{Open Data Group}\\
  \affaddr{400 Lathrop Ave Suite 90}\\
  \affaddr{River Forest IL 60305}\\
 \alignauthor Steve Vejcik\\
  \affaddr{Open Data Group}\\
  \affaddr{400 Lathrop Ave Suite 90}\\
  \affaddr{River Forest IL 60305}\\
}

\maketitle

\begin{abstract}
Developing data mining algorithms that are suitable for cloud
computing platforms is currently an active area of research, as
is developing cloud computing platforms appropriate for data mining.
Currently, the most common benchmark for cloud computing is the
Terasort (and related) benchmarks.  Although the Terasort Benchmark
is quite useful, it was not designed for data mining per se.  In this paper,
we introduce a benchmark called \malstone  that is specifically
designed to measure the performance of cloud computing middleware
that supports the type of data intensive computing common
when building data mining models.  We also introduce \malgenns, which
is a utility for generating data on clouds that can be used with
\malstonens.  
\end{abstract}

\section{Introduction}

Clouds based on the Hadoop system and associated Apache systems, such
as Hbase, Apache Pig, Hive and ZooKeeper, have proved effective for
processing large scale data for data mining and related applications
over racks of commodity computers \cite{Hadoop:2010}. This type of
architecture is sometimes called a large data cloud
\cite{Grossman:BDE09}, and was popularized in a series of Google
technical reports that described the Google File System (GFS)
\cite{Ghemawat:2003}, MapReduce \cite{Dean:2004}, and BigTable
\cite{Chang:2006}. In a large data cloud, the data is stored over many
loosely coupled distributed disks, such as you would find in racks of
commodity computers. A common architecture is for the computers in
each rack to communicate using a switch located at the top of each
rack and for different racks to communicate using a larger switch that
connects racks within a data center.

It is an important requirement now for many industry and government
applications to evaluate the applicability and scalability of
different large data cloud architectures and systems. This can be
difficult without standardized architectures and benchmarks. In this
paper, we take a first step towards a benchmark that is designed to
measure in part the ability of a large data cloud system to prepare
data for data mining and to build statistical and data mining models.

As motivation, think of the role that the TPC Benchmarks have played
in understanding performance differences between different databases
and transaction processing systems. Currently, there are no similar
benchmarks for comparing two large data clouds that support building
analytic models on large datasets. In this paper, we take a first in
this direction by introducing a benchmark called {\em \malstonens}. We
also describe the implementation of a data generator for \malstone
called {\em \malgen} as well as several experimental studies using
\malstone that compare three different large data cloud middleware
stacks.

Although using clouds for data mining and data intensive computing is
an area of active research \cite{Dean:2008}, \cite{Chu:2007}, there is
no benchmark that we are aware of for understanding the impact of
different cloud middleware on the performance of a particular
algorithm. \malstone is a first step in this direction.

Another way to view \malstone is as a stylized analytic computation of
a type that is common in data intensive computing. \malstone computes
a ratio in moving window over aggregated data. We call \malstone
stylized since it is typical of the type of derived attributes or
features that are computed as part of the modeling process. For data
small enough to fit in memory, in a disk, or in a network attached
storage system, it is straightforward to compute the \malstone
statistic. On the other hand, if the log data is so large that it
requires large numbers of disks to manage it, as is the case in a
large data cloud, then computing something
as simple as this ratio can be computationally challenging. For
example, if the data spans 100 disks, then the computation cannot be
done easily with any of the databases that are common today. On the
other hand, if the data fits into a database, then this statistic can
be computed easily using a few lines of SQL.

The open source \malgen code to generate
data for \malstone and a technical report describing some illustrative
implementations of \malstone is available from 
\begin{center}
malgen.googlecode.com.
\end{center}

\malstone was designed to give some insight into different large data
cloud systems. It was not designed to compare a large data cloud
system (in which data is typically stored over multiple distributed
disks) to a traditional database, databases and systems that utilize
proprietary hardware, or hybrid database/large data cloud systems.
Other benchmarks must be designed for this purpose. For example,
\malstone does not measure the efficiency of joins and does not take
into account the cost by TB of data managed, and similar
considerations, all of which are important when comparing these
different types of systems.

This paper is organized as follows: in Section~2 motivates the
\malstone benchmark. Section~3 describes the abstract model motivating
the statistic behind the \malstone benchmark. Section~4 describes the
benchmark. Section~5 describes \malgenns. Section~6 describes three
illustrative implementations of \malstonens and shows the sometimes
quite significant differences that can arise with different large data
cloud middleware stacks. Section~7 contains some experimental studies.
Section~8 contains some discussion. Section~9 describes related work.
Section~10 is the summary and conclusion.

This emerging application and technology paper paper makes the
following contributions:

\begin{enumerate}

\item There is currently no benchmark that we are aware for measuring
  the performance of cloud middleware designed to support building
  data mining models on large datasets. \malstone is such a
  benchmark. \malstone is defined in Section~\ref{section:benchmark}
  and is useful for quantifying differences in system architectures
  and in quantifying their scalability. See
  Tables~\ref{table:malstoneA} and \ref{table:malstoneB} for some
  examples.

\item There are currently very few data generators that we are aware
  for generating data records that can be used for testing data mining
  algorithms designed for cloud computing platforms. \malgen is such
  a data generator. \malgen is described in
  Section~\ref{section:malgen}.

\item Through three experimental studies using \malstonens, we have
  shown that are substantial differences ($\approx 20\times$)
  between different cloud computing platforms designed to support
  building data mining models on very large datasets. This is
  discussed in Section~\ref{section:experiments}.

\item The abstraction described in Section~\ref{section:spm} covers a
  number of interesting examples as summarized in
  Table~\ref{table:example}. Viewing these types of problems from this
  point of view has not received a lot attention in the data mining
  literature to date, but represents an interesting class of problems
  that occur not infrequently. As the implementations demonstrate, the
  type of log files that these types of problems produce can be
  analyzed easily using MapReduce style parallel programming
  frameworks.

\end{enumerate}

\section{Motivating Example}
We introduce \malstone with a simple motivating example. Consider
visitors to web sites. As described in the paper ``The Ghost in the
Browser'' by Provos et. al. \cite{Provos:2007}, approximately 10\% of
web pages have exploits installed that can infect certain computers
when users visit the web pages. Sometimes these are called ``drive-by
exploits.''

The \malstone benchmark assumes that there are log files that record
the date and time that users visited web pages. Assume that the log
files of visits have the following fields:
\begin{verbatim}
    Timestamp | Web Site ID | User ID
\end{verbatim}
There is a further assumption that if the computers become infected,
at perhaps a later time, then this is known. That is for each
computer, which we assume is identified by the ID of the corresponding
user, it is known whether at some later time that computer has become
compromised:
\begin{verbatim}
    User ID | Compromise Flag
\end{verbatim}

Here the Compromise field is a flag, with 1 denoting a compromise. A very simple statistic that provides some insight into whether a web page is a possible source of compromises is to compute for each web site the ratio of visits in which the computer subsequently becomes compromised to those in which the computer remains uncompromised.  This statistic is defined in Section~\ref{section:spm}.  Also, see Figure~\ref{figure:example}.

We call \malstone stylized since we do not argue that this is a useful
or effective algorithm for finding compromised sites. Rather, we point
out that if the log data is so large that it requires large numbers of
disks to manage it, then computing something as simple as this ratio
can be computationally challenging. For example, if the data spans 100
disks, then the computation cannot be done easily with any of the
databases that are common today. On the other hand, if the data fits
into a database, then this statistic can be computed easily using a
few lines of SQL.

We abstract this problem by abstracting web sites by {\em sites},
users by {\em entities}, and visits by {\em events}.  
When entity visits a {\em marked} site, it may become {\em marked}
at some time in the future.  With this
generalization, we assume that we have log files containing
events records describing an event associated with an entity and a site
and that some of these sites mark some of the entities that
are associated with them.  We assume that not all entities become
marked and that there may be a time delay in the marking.

\begin{table*}
\begin{center}
\begin{tabular}{|p{1.5in}|p{1.5in}|p{1.5in}|} \hline
{\bf Example} & {\bf Site} & {\bf Entity} \\ \hline

drive-by exploits & web site & computer identified by 
IP with browser \\ \hline

compromised login service & computer providing 
the compromised service & user providing credentials \\ \hline

\end{tabular}
\end{center}
\caption{Some examples of scenarios producing site-entity logs.  The
  problem of interest in these examples is to identify the {\em site} or {\em sites}
  that are the source of the marks assuming that we know which entities
  are marked.  The problem is difficult since: i) the site-entity log files may be
  very large; ii) there may be a background process that marks some entities
  independent of the marked sites; and iii)   not all entities that visit a marked site become marked.}
\label{table:example}
\end{table*}

\section{Sites, Entities \& Marks}
\label{section:spm}

In this section, we abstract and formalize the example described in the previous section.

\vfill\eject
\subsection{The Model}

The model we use contains abstract sites and abstract entities.  There are two types of activities: entities can {\em visit} sites and sites can {\em mark} entities.  There is a log file that records each type of activity.  The first type of log file records the times at which entities visit sites.  The second type of log file records the times at which entities become marked.  More precisely, {\em some of the entities} that visit sites became {\em marked} at {\em some time in the future} after the visit.  The second type of log file records the times at which this happens.

We use the following notation:

\begin{itemize}
\item  We usually let $e$ denote an entity and let $s$ denote a site; we also
use $e_i$ for an entity and $s_j$ for a site.

\item $A$, $B$ refer to sets of entities, $A_j$, $B_j$ refer to sets
of entities that depend upon a site  $s_j$.   

\item $S$ refers to a set of sites
and $S_i$ refers to a set of sites associated with an entity  $e_i$.

\end{itemize}

\begin{figure}
\centering
\includegraphics[scale=0.4]{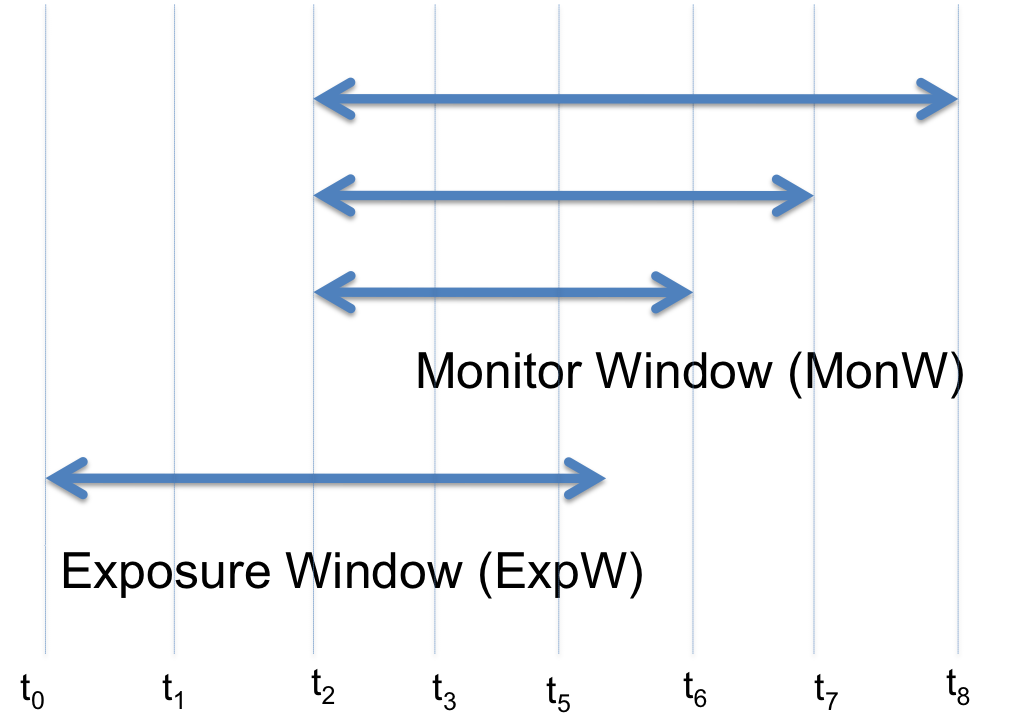}
\caption{To define the \spm\ $\rho_j$ statistic requires fixing an exposure window
and a monitor window. To define the \spm\ $\rho_{j,t}$ statistic requires fixing an exposure window
and a sequence of increasing monitor windows. 
}
\label{figure:windows}
\end{figure}

\subsection{\spm\ for Fixed Windows}

In this section, we define a statistic associated with sites, entities, and marks
called the subsequent proportion of marks or \spm.
We first define \spm-based scores for a fixed window.  In the next subsection,
we will consider moving windows.

We define the \spm\ statistic as follows:

\begin{enumerate}
\item Fix an exposure window \expwindow\  and a monitor window 
\monwindow.   See Figure~\ref{figure:windows}.
\item Fix a site $s_j$. 
\item Let $A_j$ be the set of all entities $e_i$ that: i) transact at site $s_j$ at
any time during the exposure window $\expwindow$; and ii), in the case that the entity
is marked,  the transaction occurs before the entity is marked.  
\item Let $B_j$ be the set of all entities $e_i \in A_j$ that become marked
at any time in the monitor window \monwindow.   
\end{enumerate}
\begin{definition}
Define the {\em subsequent proportion of marks} $\rho_j$ by:
$$\rho_j = \frac{ |B_j| }{| A_j |}.  $$
\end{definition}

\begin{figure*}
\centering
\includegraphics[scale=0.4]{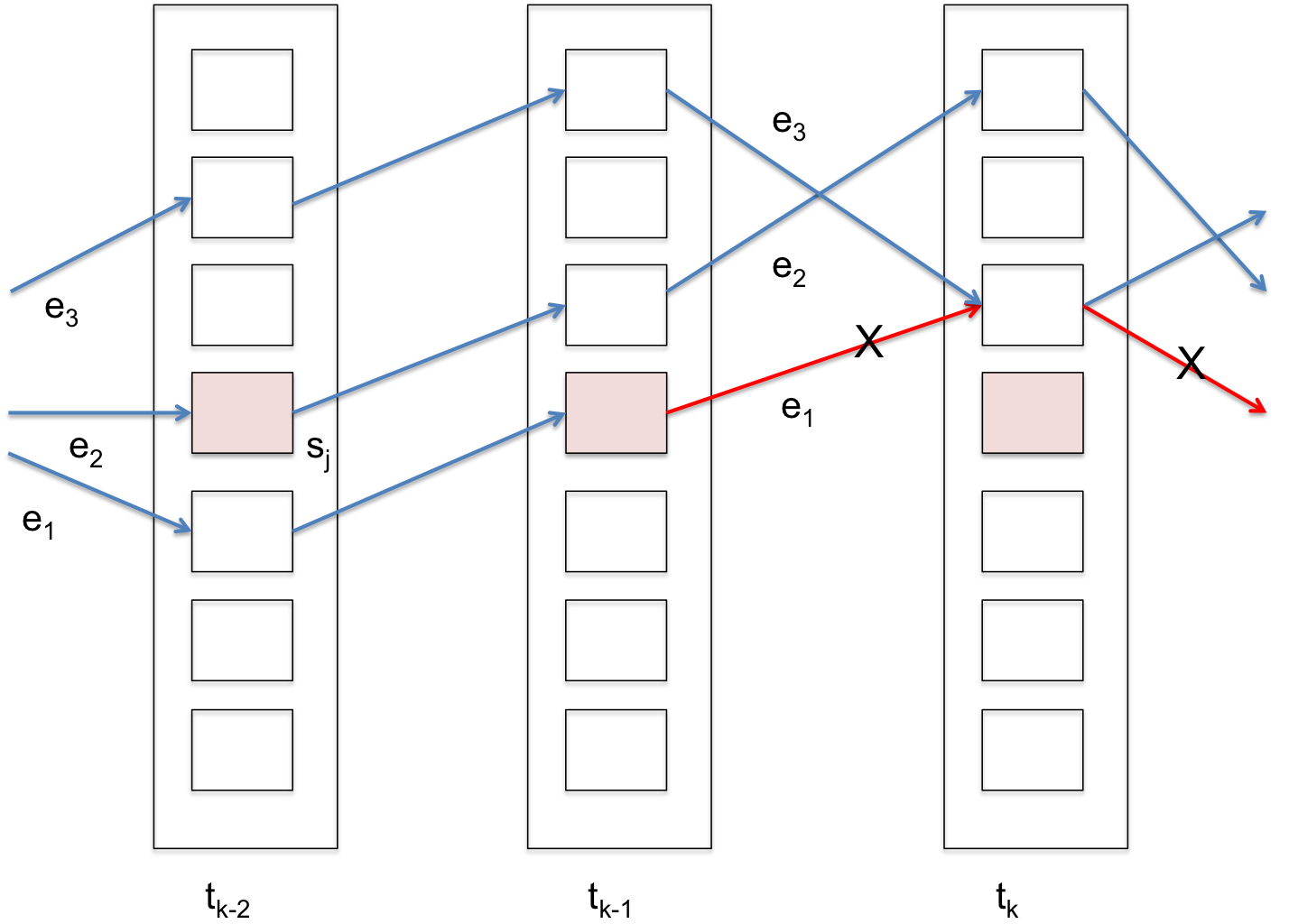}
\caption{The diagram shows an example of how the \spm\ statistic
is computed.   Here sites $s_j$ are represented by small
  rectangles and marked sites are represented by shaded
  rectangles.  Specifically, for each site $s_j$ at time $t_k$,
  \malstone B collects all the transactions (represented by arrows)
  that are associated with the site at time $t_k$ or earlier. Notice
  there are no transactions associated with $s_j$ at time $t_k$, but
  that there are two transactions associated with the site at earlier
  times $t_{k-1}$ and $t_{k-2}$.  Entity $e_2$ was associated with the
  site at $t_{k-2}$ and entity $e_1$ at time $t_{k-1}$.  Entity $e_1$
  became marked at the site $s_j$ at time $s_{k-1}$ (represented
  by red entity arrow with an ``X''). Therefore $\frac{1}{2}$ of the
  transactions are marked for site $s_j$ with respect to the
  window $(t_{k-2} , t_{k-1} , t_k)$.}
\label{figure:example}
\end{figure*}

We close this section with some remarks:
\begin{itemize}

\item Note that $B_j \subseteq A_j$.

\item It is important to note
that $A_j$ depends upon the exposure window, and $B_j$ depends upon the
monitor window, and through the relation $B_j \subseteq A_j$ upon the exposure window.

\item Note that the \monwindow\ may: 1) start after the \expwindow, 2) before
the \expwindow, or 3) include the entire data available.

\item As an example, in Figure~\ref{figure:example}, 
for time $d_k$, there are no events for
site $s_j$, but for the window starting at time $d_k$, extending backward to
time $d_{k-2}$ (time zero in this example), the statistic is 
$(1 + 0 + 0) / (1 + 1 + 0) = \frac{1}{2}$.

\end{itemize}

\subsection{\spm\ for Moving Windows}

In general, we have a sequence of monitor windows  
$$\monwindow_{t_1},  \monwindow_{t_2}, \monwindow_{t_3}, \ldots $$
depending upon time $t$.  For example, the sequence may all have a common
start time, but the end time increases by a week for each monitor window
in the sequence.  See Figure~\ref{figure:windows}.

Using this sequence of monitor windows, we can define a sequence
$$\rho_{j,t} = \frac{ |B_{j,t}| }{| A_j |},  $$
where $B_{j,t}$ is the set of entities $e_j\in A_j$ that become marked
at any time during the monitor window $\monwindow_t$.

\section{\malstone A \& B}
\label{section:benchmark}
In this section, we define the MalStone A and B Benchmarks.

Assume that we have a collection of log files.  For simplicity, we assume
that the log files that describe visits of entities to sites has been joined
to the log file that describes which entities are marked (and when).
With this assumption, log files contain the following fields:
\begin{verbatim}
  Event ID | Timestamp | Site ID | 
      Entity ID | Mark Flag |
\end{verbatim}
We interpret these as recording the fact that at the time indicted by the timestamp,
the entity with the entity ID visited the site with the Site ID.  The Mark Flag indicates
whether at the time of visit the entity was marked.  

\medbreak
{\bf Remark.}  
It is important to note that if the Mark Flag is 1 indicating the entity is marked, we do not necessarily
know that the site identified by the Site ID marked the entity.  Instead, all that
we know is that {\em either the site, or any site that the entity has visited in the past
during the exposure window} has marked the entity.   

It is for this reason, that the statistic is called the subsequent proportion of marks.

\malstone A computes $\rho_j$ for all sites $j$ in the log files.  \malstone B computes
$\rho_{j,t}$ for sites $j$ in the log files and for a sequence of moving windows 
that begin at time $t_0$ and end at time $t$ equal to:
$$ t_1 < t_2 < t_3 < \ldots $$

MalStone records are 100 byte records, with a fixed width fields.
Both \malstone A and B use 1 year's worth of data.  \malstone A uses a
single window for the entire year, while \malstone B uses a window that
begins at the beginning of the year and ends at week 1, week 2, $\ldots$,
week 52.

\begin{table}
\begin{tabular}{|l|l|l|l|} \hline
{\bf Benchmark} & {\bf Statistic} & {\bf \#  records} & {\bf Data Size} \\ \hline
MalStone A-10 & $\rho_j$ & 10 billion & 1 TB \\ \hline
MalStone A-100 & $\rho_j$ & 100 billion & 10 TB \\ \hline
MalStone A-1000 & $\rho_j$ & 1 trillion & 100 TB \\ \hline
MalStone B-10 & $\rho_{j,t}$ & 10 billion & 1 TB \\ \hline
MalStone B-100 & $\rho_{j,t}$ & 100 billion & 10 TB \\ \hline
MalStone B-1000 & $\rho_{j,t}$ & 1 trillion & 100 TB \\ \hline
\end{tabular}
\caption{The MalStone benchmarks use 100 byte records, with a fixed
field width. }
\label{malstone-table}
\end{table}

\vfill
\section{\malgen }
\label{section:malgen}
As mentioned above, we have developed an open source program called
\malgen for generating site-entity log files for all the nodes in a
cluster.  \malgen uses a power law distribution to model the number of
entities associated with a site.  Most sites have few entities
associated with them, while a few sites have a large number of
entities.  This is the case, for example, with web sites: most sites
have only a few visitors, a few sites have a lot of visitors, and a
power law distribution is often used to model this distribution.

MalGen are 100 bytes in size with five fixed width fields:
\begin{verbatim}
  Event ID | Timestamp | Site ID | 
     Entity ID | Mark 
\end{verbatim}
The following is a description of each field:

\begin{itemize}
\item Event ID ---  The Event ID consists of an ID for each record that is sequential and unique when restricted to a single node followed by a hash of the hostname to create a globally unique Event  ID.

\item Timestamp --- The date and time of the event. This is a
  uniformly distributed random value over a user-specified number of
  days.  The default is to generate data distributed over a period of
  one year.

\item Site ID --- This is the ID of the site associated with the event. 

\item Entity ID -- This is the ID of the entity associated with the  event.

\item Mark --- The field is either 0 or 1 and indicates whether
the entity is marked at the time.  Note that, as discussed above, the fact that the mark is 1 does 
not indicate that the site with Site ID is responsible for the mark, simply that at some time
prior to the Timestamp. the entity visited some marked site.

\end{itemize}

%
%

Site-entity log files are generated by \malgen in several steps. 
Several steps are used since the \malstone \spm statistic
requires aggregating data from different distributed nodes and
computing a statistic that satisfies both certain statistical properties
and certain consistency requirements.  The
first step generates certain seed information about the marked sites and scatters
this information to all the nodes in the large data cloud.   The
algorithm is designed to keep certain information required for the
first step  in memory in order to improve the overall speed of \malgen.  
The subsequent steps are done independently by each of the nodes.
We report on the memory utilization of the first step of \malgen below since keeping
the memory utilization relatively low is important so that enough seed
information is available for generating the 10 billion, 100 billion
and 1 trillion records that \malgen requires.

In the first step, \malgen generates events associated with marked sites.
For each marked site, a random date is generated. 
For a particular site, the number of events is randomly
generated using a power law distribution and a set of entity IDs is
randomly generated from the pool of available entity IDs. The power
law distribution is constructed so that most sites are associated with
a relatively few number of entity IDs (a few hundred a day), but with
a long tail so that there are a small number of sites with a very
large number of events.  The Entity IDs are sampled until the
number of events for each site is complete.

For the marked sites, a visit by an entity subjects
the entity to a probability (e.g. 70\%) of being marked.
If an entity is marked, it is tagged as being such with a
timestamp that occurs after a delay period (e.g. one week)
If an entity is already marked when it 
visits a marked site again, it is subject to mark if the date of
the current event precedes that of the event that marked it. 
In this case, the date-time of the mark is updated
accordingly. 

The initial seeding and the generation of marked entities is done
on a single node.  This information is then pushed out to all the nodes
in the large data cloud and each local node then generates records for entities
that are not marked.  Table~\ref{table:malgen-times} shows the times
required to generate 2 billion, 6 billion and 10 billion events in this
way on a 20 node cloud.

The time required for seeding the process is in the table below:
\begin{table}
\begin{center}
\begin{tabular}{|l|l|l|} \hline
Records/node & RAM & Time \\ \hline
100 M & 16 GB & 60 min \\ \hline
300 M & 16 GB & 142 min \\ \hline
500 M & 16 GB & 190 min \\ \hline
\end{tabular}

\medbreak
\medbreak

\begin{tabular}{|l|l|l|l|} \hline
Records/node & Total Records & RAM & Time \\ \hline
100 M & 2 B & 4 GB & 54 min \\ \hline
300 M & 6 B & 4 GB & 157 min \\ \hline
500 M & 10 B & 4 GB & 275 min \\ \hline
\end{tabular}

\caption{The first table shows the time required in minutes for
  \malgen to seed the data generation and to generate the marked
  entities.  This was run on a head node with 16 GB of memory.  The
  second table shows the time required to copy the required data from
  the head node to each of 20 local nodes and for the local nodes to
  generate all the required unmarked events.  For example, the time
  required to generate 10 billion events distributed over 20 nodes is
  $190+275=465$ minutes. }
\label{table:malgen-times}
\end{center}
\end{table}

After all event histories for the requested number of marked
sites are complete, subsequent sites are assumed to be unmarked
with no possibility of being marked.  This is the third step,
which uses the same process that was just described to construct
visits for non-marked sites.

We close this section with two remarks about \malgenns.

\malgen is designed to generate large datasets that span all the nodes
in a cluster.  To create consistent data in parallel over all the
nodes in the cluster, \malgenns: i) generates the data describing all
marked sites on one machine in the cluster; ii) scatters the
information to all the other nodes in the cluster; iii) all the other
nodes in the cluster then generate the data for all the unmarked
sites.

\begin{figure}
\centering
\includegraphics[scale=0.6]{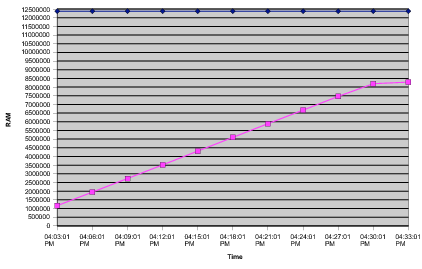}
\caption{The memory usage of \malgen as it generated approximately
500,000,000 log records for visits to approximately 120,000 different
sites, where the number of visits to a web site follows a power law.
Approximately 8.5 GB of an available 12 GB of memory were used during
the 30 minute generation of data.  The bottom line shows the memory used,
while the top line shows the available memory.}
\label{figure:memory}
\end{figure}

By keeping information about sites in memory (vs on disk), \malgen can
improve its performance.  The default parameters for \malgen can
generate approximately 500,000,000 events for approximately
120,000 sites in about 30 minutes using a Dell 1435 2.0GHz dual-core
AMD Opteron 2212 processor and 16 GB of memory.
Figure~\ref{figure:memory} contains a graph showing \malgenns's memory
usage.

\section{Three Implementations}

We have implemented the \malstone A and B benchmarks in three different
ways: 
\begin{enumerate}
\item using the Hadoop Distributed File System (HDFS) \cite{Borthakur:2007}
and Hadoop's implementation of MapReduce \cite{Hadoop:2010}; 
\item using HDFS, Hadoop Streams \cite{Hadoop:2010} and coding \malstone 
A and B in Python; and
\item 3) using the Sector Distributed File System
and Sphere User Defined Functions \cite{Grossman:PTRSA09}.
\end{enumerate}

\subsection{Hadoop HDFS and MapReduce}
\label{section-hadoop}

We used \malgen to generate data which we stored in HDFS and then
implemented \malstone B using a Mapper, Reducer and Partitioner as follows:

\begin{itemize}

\item Mapper --- Reads the records and groups them using the Site ID
as the key.  The corresponding value is the timestamp and the mark flag.

\item Reducer --- For each key, the Reducer tracks the total number of
  events seen with the mark flag equal to one and the total number of
  events and stores them by the date.  All saved values are then
  processed in order by the date.

\item Partitioner --- The Site Id is taken modulo the number of reducers.

\end{itemize}

Each record is parsed and then transformed into a key-value pair.  The
key is the Site Id.  The value is the Flag and the bucket the time
stamp is put in.  The time stamp in each record can be bucketed
arbitrarily.  \malstone B requires that the statistic be computed for
each week; we used the ISO week number (www.iso.org) for convenience.

The operation performed on each group of data is to count the number
of events and the number of events with the mark equal to one
each time $t$.  For each site $s$, this is stored in a Java Collections
Map using $t$ as the key.

When all records for a site id are processed, the stored values are
accessed in chronological order and running totals computed.

The output is the Site ID (key) $j$ and a list of the times $t$ and 
associated SPM statistics $\rho_{j,t}$

\subsection{HDFS, Hadoop Streams and Python}

The second implementation used Hadoop Streams \cite{Hadoop:2010} and 
Python.  The mapper method reads the records from Standard Input
and sends the mapped data to Standard Output.  The same key and value structure
as described in Section~\ref{section-hadoop} is used.

The reducer reads the mapped data from Standard Input and for each
Site ID, stores the aggregated number of events seen and those seen
with the mark equal to 1 in a Python dictionary keyed by the time $t$.

When all the records for a site id are processed, the stored values
are accessed in chronological order and running totals computed.

The output is the Site ID (key) and a list of the times $t$'s and the
associated \spm  statistic $\rho_{j,t}$ is sent to Standard Output (value).

\subsection{Sector and Sphere UDFs}

Sector provides two methods for implementing processing  \cite{Grossman:PTRSA09}: 

\begin{itemize}

\item Using indexed data --- when using indexed data each input data
  file has an accompanying index file containing the offsets of each
  record in the data file. This index allows Sector to segment the
  data during processing.

\item Using non-indexed data --- when using non-indexed data, the
  processing code must manually segment the input the input data during processing.

\end{itemize}

Using non-indexed data requires somewhat more code to implement, but
seems to improve processing time.  \malstone B was implemented using
non-indexed data.

The \malstone B code was implemented in two stages:

\begin{itemize}

\item In the first stage, each record in the input data is read, assigned to a bucket based upon the site ID, and each bucket file is written to disk.  After this stage is completed, all records for a particular site will be in a single file.

\item In the second stage, for each site, for each site $j$ the cardinality of the sets $A_j$ and $B_j$ is computed.  After all the records for a site $j$ have been processed, the resulting record is saved to a file.

\end{itemize}

\section{Experimental Studies}
\label{section:experiments}

\subsection{Testbed}

The experimental studies used a rack of 30 Dell 1435 computers.
Each computer had 12 GB memory, 1TB disk, and a 
2.0GHz dual dual-core AMD Opteron 2212.  Each computer 
had a 1 Gb/s network interface cards and were networked together
with a Cisco 3750E switch.

\subsection{\malstone Benchmarks}

For the experimental studies reported below, we used 20 nodes. Each
node was populated with 500 million records using \malgen for a total
of 10 billion records.  Each record was 100 bytes for a total of 1 TB
of data.  The results are reported in Tables~\ref{table:malstoneA} and
\ref{table:malstoneB}.

Note that as measured by these benchmarks, storing the data using HDFS and 
implementing the benchmark using Hadoop streams and Python was substantially
faster than using HDFS and Hadoop's MapReduce.

Note also that managing the data using Sector and implementing the
benchmark using Spheres UDFs was about 2.5 times faster than the
Hadoop streams implementation.

\begin{table*}
\begin{center}
\begin{tabular}{|p{1.0in}|p{1.0in}|p{1.0in}|p{1.0in}|} \hline
& {\bf Hadoop HDFS with Streams \& Python} & 
{\bf Hadoop HDFS with MapReduce} & {\bf Sector with 
Sphere UDFs} \\ \hline
Run 1 & 
	82m 21s & 
	458m 7s & 
	33m 44s \\ \hline
Run 2 & 
	90m 31s & 
	450m 21s & 
	33m 26s \\ \hline
Run 3 & 
	89m 35s & 
	454m 12s & 
	33m 51s \\ \hline
Average	 & 
        87m 29s & 
	454m 13s & 
	33m 40s \\ \hline
\end{tabular}
\end{center}
\caption{This table summarizes an experimental study
running \malstone A on 20 nodes.  Each node had 500 million 100-byte
\malstone records.  The tests used version 0.18.3 of Hadoop
and version 1.20 of Sector.}
\label{table:malstoneA}
\end{table*}

\begin{table*}
\begin{center}
\begin{tabular}{|p{1.0in}|p{1.0in}|p{1.0in}|p{1.0in}|} \hline
& {\bf Hadoop HDFS with Streams \& Python} & 
{\bf Hadoop HDFS with MapReduce} & {\bf Sector with 
Sphere UDFs} \\ \hline
Run 1	& 
        144m 10s &
	799m 0s &
	43m 57s  \\ \hline
Run 2  &
	146m 23s  &
	861m 40s &
	43m 52s \\ \hline
Run 3 &
	137m 4s &
	861m 51s &
	43m 24s  \\ \hline
Average	 &
        142m 32s &
	840m 50s  &
	43m 44s   \\ \hline
\end{tabular}
\end{center}
\caption{This table summarizes running \malstone B on 20 nodes.
        Each node had 500 million 100-byte
        \malstone records.  The tests used version 0.18.3 of Hadoop
        and version 1.20 of Sector.}
\label{table:malstoneB}
\end{table*}

\vfill
\section{Discussion}

One of the pleasant surprises is the power of Hadoop streams, which
does not require the MapReduce framework. Hadoop is now a relatively
mature distributed file system that can scale to over a thousand nodes
and manage petabytes of data. As Tables~\ref{table:malstoneA} and
\ref{table:malstoneB} show, Python programs can be invoked by Hadoop
streams and be used to efficiently process large data sets without the
MapReduce framework and this approach can be faster when computing
certain statistics (such as $\rho_{j,t}$) than performing the
computation using MapReduce. We stress that this is a positive outcome
and simply shows (as is obvious in hindsight) that certain statistical
qualities can be computed more efficiently directly with Python over
the data managed by the HDFS than by using MapReduce and the HDFS.

Another pleasant surprise is that once we abstracted the \malstone
statistic as the Site-Entity-Mark Model, we found that other
applications could also be modeled in this way, as
Table~\ref{table:example} shows.

In practice, once the \malstone \spm\ statistic is computed,
relatively effective statistical models can be computed by looking
for changes over time $t$ in the $\rho_{j,t}$ statistic using CUSUM,
GLR and related change detection models
\cite{Poor:QuickestDetection2008}.  Although outside the scope of this
paper, if segmented models are used for each site $j$, the Reducer in MapReduce 
can be used to organize the computation so that each node in a large
data cloud contains all the data required to build a change detection
model for a site $j$ \cite{Grossman:Sawmill}.

\section{Related Work}

The CloudStone Benchmark \cite{Sobel:2008} is a first step towards a
benchmark for clouds designed to support Web 2.0 type applications. In
this note, we describe the \malstone Benchmark, which is a first step
towards a benchmark for clouds, such as Hadoop and Sector, designed to
support data intensive computing.

One of the motivations for choosing 10 billion 100-byte records is
that the TeraSort Benchmark \cite{Gray:TeraSort} (sometimes called the
Terabyte Sort Benchmark) also uses 10 billion 100-byte records.

We note that in 2008, Hadoop became the first open source program to hold the
record for the TeraSort Benchmark. It was able to sort 1 TB of data
using using 910 nodes in 209 seconds, breaking the previous record of
297 seconds.  Hadoop set a new record in 2009 by sorting 100 TB of data
at 0.578 TB/minute using 3800 nodes. 

The TeraSort Benchmark is now deprecated and has been
replaced by the Minute Sort Benchmark. Currently, 1 TB of data can be
sorted in about a minute given the right software and sufficient
hardware.  

The paper by Provos et. al. \cite{Provos:2007} describes a system for
detecting drive-by malware that uses MapReduce.  Specifically,
MapReduce is used to extract links from a large collection of crawled
web pages.  These links are then analyzed using heuristics to identify
a relatively small number of suspect web sites.  These suspect web
sites are then tested using Internet Explorer to retrieve web pages in
a virtual machine that is instrumented.  This allows those web sites
resulting in drive-by infections to be directly monitored.  In
contrast, the work described in this paper is quite different.  The
work here uses Hadoop and MapReduce to compute the \malstone statistic
from a collection of log files generated by \malgen in one of the
illustrative implementations of \malstonens.

The paper \cite{Chu:2007} describes how several standard data mining
algorithms can be implemented using MapReduce, but this paper does not
describe a computation similar to the \malstone statistic.

\section{Summary}

In this paper, we have introduced a benchmark called \malstone for
measuring the performance of cloud middleware designed for data mining
and data intensive computing.  Currently, gaining access to large
amounts of nonproprietary data to use for benchmarking cloud
middleware can be challenging.  For this reason, we have developed an
application called \malgen that is designed to generate synthetic
log-entity files that can be used by \malstonens.  We have used \malgen
to generate tens of billions of events on clouds with over 100 nodes.

The \malstone benchmark computes a statistic that is a stylized analytic on log files consisting of records of visits by entities to sites.  Sometimes, after these visits, entities become marked at some time in the future.  Note that this analytic is related to identifying the {\em sites} that are the sources of the marks, not the marked {\em entities} themselves.

As Tables~\ref{table:malstoneA} and \ref{table:malstoneB} show, there
can be substantial differences in performance, depending upon which
cloud middleware is used to compute the \malstone statistic.

\section{Availability}

\malgen is open source and available from:
\begin{center}
malgen.googlecode.com.
\end{center}
The current version of \malgen is 0.9.

\end{document}